# Microdroplet Approach for Measuring Aqueous Solubility and Nucleation Kinetics of a Metastable Polymorph: The case of KDP Phase IV

Ruel Cedeno[1], Romain Grossier[1], Nadine Candoni[1], Stéphane Veesler[1]*

[1]CNRS, Aix-Marseille Université, CINaM (Centre Interdisciplinaire de Nanosciences de Marseille), Campus de Luminy, Case 913, F-13288 Marseille Cedex 09, France

**TOC Graphic**

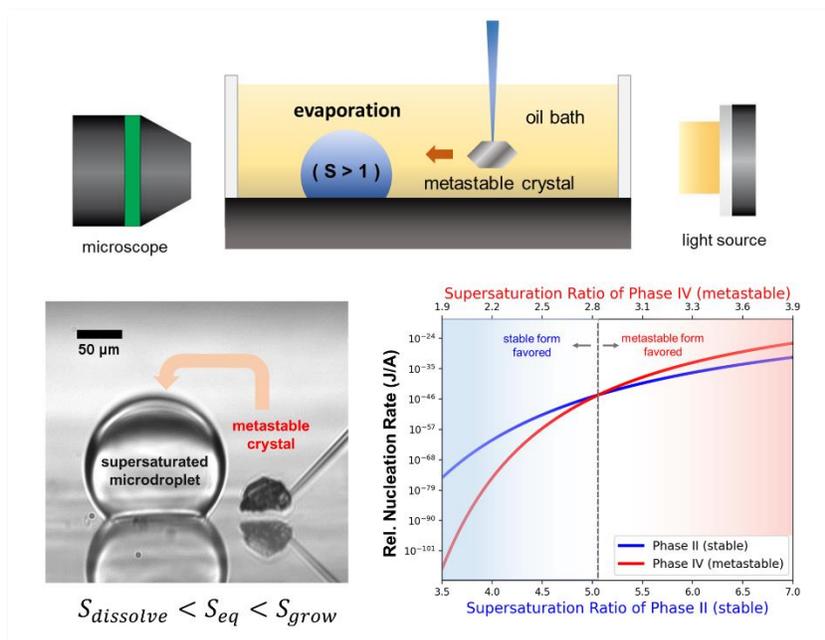

**Abstract**

Solubility and interfacial energy are two fundamental parameters underlying the competitive nucleation of polymorphs. However, solubility measurement of metastable phases comes with a risk of solvent-mediated transformations which can render the results unreliable. In this work, we present a rapid microfluidic technique for measuring aqueous solubility of the metastable form using KDP Phase IV as a model system. This bracketing approach involves analyzing the dissolution behavior of crystals in contact with supersaturated microdroplets generated via evaporation. Then, with the help of our recently developed nucleation time measurement technique, together with Mersmann calculation of interfacial energies from solubilities, we were able to access the interfacial energies of both metastable and stable phases. To gain further insights into the observed nucleation behavior, we employed the Classical Nucleation Theory (CNT) to model the competition of polymorphs using our measured solubility and calculated interfacial energies. The results show that the stable form is favored at lower supersaturation while the metastable form is favored at higher supersaturation, in good agreement with our observations and experimental reports in the literature. Overall, our microfluidic approach allows access to unprecedentedly deep levels of supersaturation and reveals an interesting interplay between thermodynamics and kinetics in polymorphic nucleation. The experimental methods and insights presented herein can be of great interest, notably in the mineral processing and pharmaceutical industry.

**INTRODUCTION**

The solubility of metastable polymorphs with respect to the stable form is an important information required in quantifying the kinetics of competitive polymorphic nucleation. However, obtaining the solubility of the metastable phases can be challenging due to the risk of solvent-mediated transformation during measurement. This is the case for potassium dihydrogen phosphate ($KH_2PO_4$ or KDP), one of the most important optoelectronic materials.[1] While its tetragonal polymorph (Phase II, $I\bar{4}2d$) is the most stable at ambient conditions, KDP can crystallize in several polymorphic forms depending on temperature and pressure.[2] Recently, the high pressure monoclinic polymorph (Phase IV, $P2_1c$) that is usually accessible at 1.6 GPa has been crystallized at ambient conditions. This was achieved by carefully eliminating impurities and potential heterogeneous nucleation sites such as in specially-treated glass[1] or in levitating droplets.[3] Moreover, KDP Phase IV has been shown to have interesting optical properties.[1] However, due to the lack of solubility data for Phase IV, there has been no quantitative studies on its nucleation barrier in comparison with the stable tetragonal Phase II. Indeed, the solubility of the metastable polymorph in water is difficult to measure using conventional methods[4] because it readily converts to the stable tetragonal Phase II form upon contact with moisture.[5] As a result, the interfacial energies governing the competitive nucleation kinetics of the metastable and stable polymorphs are currently lacking.

In this work, we report a rapid microdroplet approach to estimate the ambient aqueous solubility of the metastable Phase IV KDP. This involves analyzing the dissolution behavior of crystals of the metastable phase in contact with supersaturated microdroplets, with respect to the stable phase, generated via evaporation. Moreover, with the help of our measured solubility and classical nucleation theory, we calculated its interfacial energy by employing our recently-developed microdroplet-based nucleation measurement technique.[6] Furthermore, upon applying the empirical correlation of Mersmann[7] between solubility and interfacial energy, we analyze the interplay of thermodynamics and kinetics in the competitive nucleation of the metastable and stable polymorph. Our investigation on KDP polymorphs reveal interesting insights into the competitive nucleation of stable and metastable crystals.

**MATERIALS AND METHODS**

**Solubility Measurement.** The schematic diagram of the setup is shown in **Figure 1a**. Given that the solubility of the metastable polymorph is higher than that of the stable polymorph, the solution that would equilibrate with the metastable polymorph must be supersaturated with respect to the stable polymorph. To generate supersaturated solutions while avoiding the nucleation of the stable polymorph, we took advantage of the long nucleation times in microdroplets (i.e. nucleation time is inversely proportional to volume). To ensure that the evaporation rate is slow enough to reach quasi-equilibrium, we submerged the microdroplets in an oil bath (0.8 mm layer of PDMS 10 cSt). For perspective, sessile microdroplets (2-3 nL) in air would evaporate within few seconds while those under oil would take several hours (see Figure S1 of SI). As shown in our previous work,[8] such a system maintains a uniform concentration distribution in the microdroplet, i.e. Peclet number << 1. As the initially saturated microdroplet evaporates, the supersaturation $S$ (i.e. $c/c_{sat}$ where $c_{sat}$ is the saturation concentration) slowly increases. Consequently, different supersaturation levels can be achieved by varying the evaporation time. The supersaturation ratio $S$ is calculated from the ratio of the volume with respect to the initial volume, measured from the lateral optical images. The metastable crystal is then brought in contact with the supersaturated microdroplet using a microcapillary. For any given supersaturation $S$, if the crystal dissolves, solubility > $S_{dissolve}$ ; conversely, if the crystal grows, then the solubility < $S_{grow}$. The procedure is done for several incremental supersaturation levels until a reasonable precision is achieved on the lower and upper bound of solubility, i.e. $S_{dissolve}$ < solubility < $S_{grow}$.

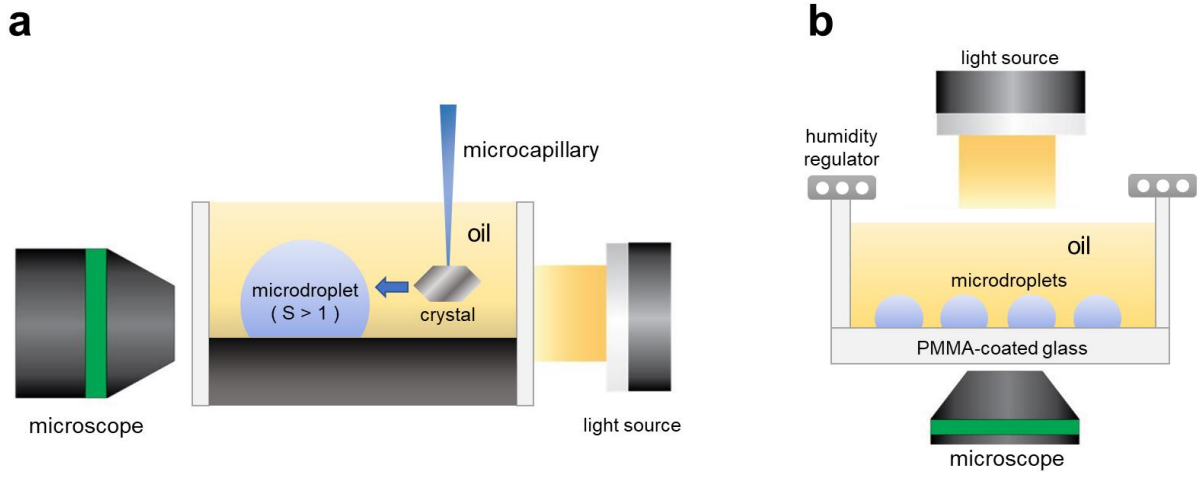

**Figure 1.** (a) Schematic diagram of solubility measurement. Supersaturation is achieved via evaporation starting from initially saturated microdroplet. The dissolution behavior of the metastable crystal when brought in contact with the supersaturated microdroplet is probed by lateral imaging (b) Schematic diagram of the interfacial energy measurement in which the probability distribution of nucleation times is obtained from hundreds of droplets via image analysis.[9] (Figure adapted from Ref.[10])

**Nucleation time and interfacial energy measurement.** The schematic diagram of the setup is shown in **Figure 1b** based on a previously reported experiment protocol in Ref[6]. However, unlike NaCl that dissolves upon exposure to humidity (RH > 75%), KDP does not absorb enough moisture to undergo deliquescence, preventing the use of the RH cycling technique. Thus, we modify the procedure by generating initially saturated microdroplets (i.e. saturation time = 0). This is done at RH close to 100% to minimize possible evaporation during microdroplet generation. Once the desired number of microdroplets are generated, the RH is lowered to 10%, causing the droplets to evaporate and eventually nucleate. Using our tailor-made evaporation model[10], the supersaturation ratio of the microdroplets can be obtained as a function of time. Then, we employed the inhomogeneous Poisson distribution coupled with classical nucleation theory to extract the interfacial energy from the probability distribution of nucleation times which can be written as

$$P(t) = 1 - \exp\left[-\int_{t_{sat}}^{t_{nuc}} J(t)V(t)dt\right] \qquad (1)$$

with *P(t)* as the fraction of microdroplets nucleated after time *t*, *V(t)* droplet volume at time *t*, $t_{sat}$ time at which the microdroplet becomes saturated and $t_{nuc}$ nucleation time.

For the primary nucleation rate *J(t)*, we used the classical nucleation theory (CNT) as

$$J(t) = A \exp\left[-\frac{16\pi}{3} \frac{\gamma_{eff}^3}{\rho_s^2 (k_b T)^3 \ln^2 S(t)}\right] \qquad (2)$$

where $A$ is the pre-exponential factor, $\gamma_{eff}$ is the effective interfacial energy, $\rho_s$ is the number density of formula units in the solid (2.27 × $10^{28}$ m$^{-3}$ for NaCl), $k_b T$ is the thermal energy, $S(t)$ is the supersaturation expressed as the ratio of concentrations (concentration at nucleation / concentration at saturation) at nucleation at time $t$ (more details in Ref[8]).

**RESULTS AND DISCUSSION**

Initially, we have observed that the crystal nucleated from microdroplet dissolves in the presence of the stable crystal (see Video 1 of SI), suggesting the formation of a more soluble metastable phase. To identify this phase, we characterized the crystals using powder X-ray diffraction (PXRD) and compared it with the simulated patterns of known polymorphs. The result is shown in **Figure 2a** which indicates a match between our crystals and the KDP Phase IV of Ren et al.[1] While the Phase IV crystallized at high pressure (1.6 GPa) has been observed to transform to Phase I upon pressure release,[2] our Phase IV from microdroplets remain stable at ambient pressures as evidenced by the absence of Phase I peaks. This is also in agreement with Ren et al.[1] The SEM image of the Phase IV shown in **Figure 2b** also reveals a crystal habit that is different from what is expected for the stable tetragonal phase II polymorph.[11]

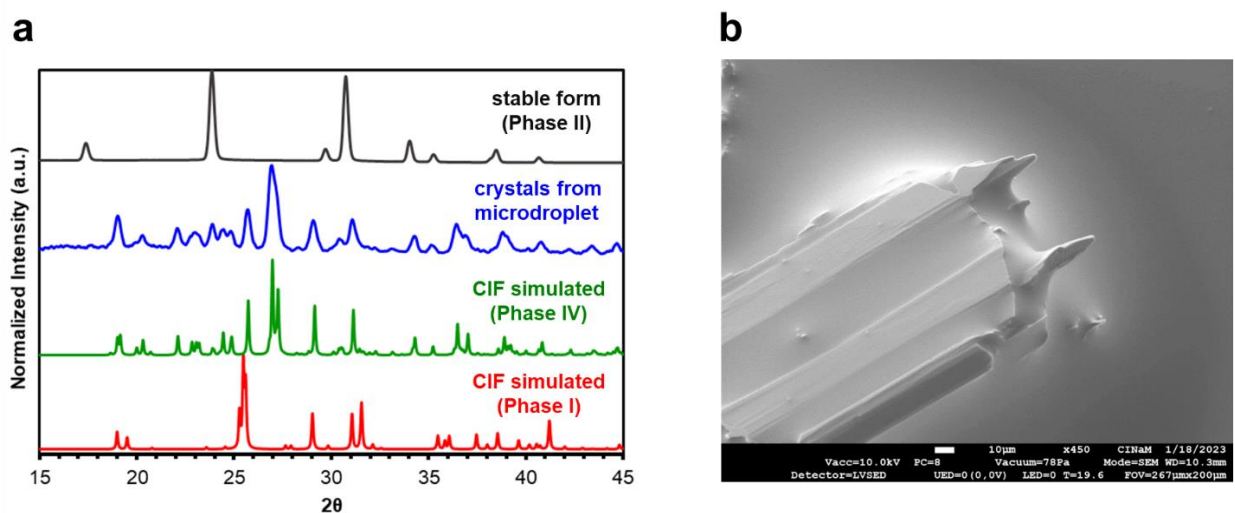

**Figure 2.** (a) Powder X-ray diffraction pattern of the commercially available stable tetragonal polymorph (black), crystals obtained from microdroplet (blue), and the CIF simulated patterns of Phase IV (green) and Phase I (red) polymorphs (b) SEM image of the crystal from microdroplet

Having confirmed the formation of KDP Phase IV, we proceed by investigating its aqueous solubility. Note that the solubility is required in evaluating its supersaturation ratio and in turn, its associated nucleation free energy barrier. However, aqueous solubility measurement of a metastable form is technically difficult, especially in the case of KDP which has been shown to undergo transformation to the stable Phase II in the presence of moisture.[5] Consequently, the risk of polymorphic transformation during measurement would render the conventional techniques unreliable. To address this, we report the use of a simple approach to rapidly estimate the solubility of the metastable form. This approach takes advantage of microdroplets evaporating in oil wherein (1) highly supersaturated states can be achieved without premature nucleation of the stable form (2) uniform concentration distribution is achieved, i.e. Peclet number << 1, (3) slow evaporation permits sufficient crystal-solution equilibration.

To demonstrate this approach, the snapshots of the solubility measurement in shown in **Figure 3** (corresponding videos 2-4 available in SI). Notice that at supersaturation ratio $S$ = 1.5, the metastable crystal dissolved upon contact with the microdroplet. In contrast, at $S$ = 2.5, the crystal grew. This suggests that the solubility of phase IV lies between 1.5 and 2.5. We then tested several values within this range and the results are displayed in **Table 1**.

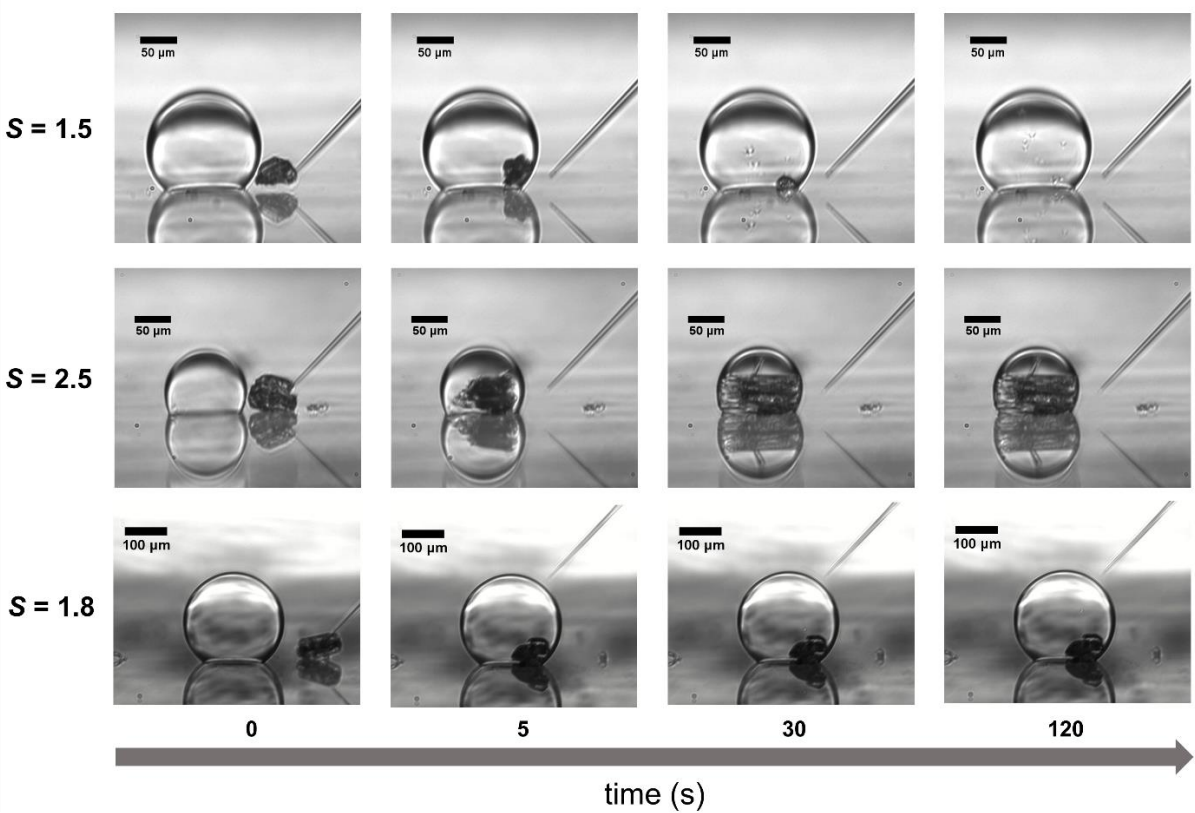

**Figure 3.** Snapshots during the solubility measurement of phase IV at selected times for different supersaturation ratios, with respect to phase II, performed at ambient conditions (25°C, 1 atm, RH=0.35).

**Table 1.** Dissolution behavior of the metastable crystal inside the microdroplet as a function of supersaturation ratio (with respect to the stable form) in the bracketing experiment.

| Supersaturation Ratio S with respect to the stable form | Outcome |
|---|---|
| 1.5 | dissolve |
| 1.6 | dissolve |
| 1.7 | dissolve |
| 1.8 | ~constant |
| 1.9 | grow |
| 2.5 | grow |

We found that at $S$=1.8, the crystal essentially undergoes neither dissolution nor growth, indicating an equilibrium. Thus, the solubility can be conservatively estimated at $S$ = 1.8 ± 0.1 (corresponds to around 45 ± 2 g KDP per 100 g water at 25°C)[12]

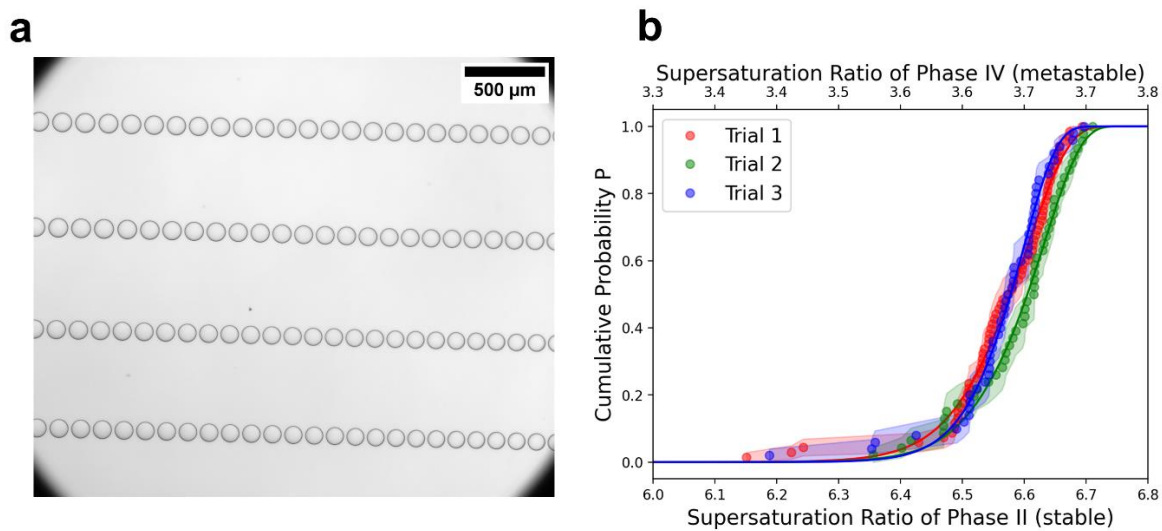

**Figure 4.** (a) Bottom-view image of sessile microdroplets (~200 pL at saturation) used in the nucleation kinetic measurement performed at 25°C, 1 bar, 10% RH. This process yields KDP Phase IV as confirmed by PXRD. (b) Cumulative probability plot (N= 68, 46, 50 for Trials 1,2,3 respectively) as a function of supersaturation ratio with respect to stable form (bottom axis) and metastable form (top axis) fitted with inhomogeneous Poisson distribution coupled with classical nucleation theory[8].

With the knowledge of the solubility of the metastable polymorph, we would be able to measure its interfacial energy from its probability distribution of nucleation times using the experimental setup in Ref[8]. The bottom-view optical image is shown in **Figure 4a** and the resulting probability plots for three trials are presented in **Figure 4b**. As part of data cleaning and pre-processing of induction time results, we applied a statistical test ($k$-sample Anderson Darling test) to identify and eliminate outliers in the dataset.[6]

We also removed the droplets that showed evidence of droplet interactions (using the image analysis protocol in Ref[6]). Then, we calculated the supersaturation ratio as a function of dimensionless induction time using a tailor-made evaporation model[10] (details in section S1 of SI). The calculated Peclet number (Pe) suggests that the droplets maintain a homogeneous concentration distribution (i.e. Pe << 1, Figure S2d of SI). It is worth noting that for aqueous KDP solutions, we achieved unprecedentedly deep levels of supersaturation (S > 6 as shown in Figure 4b) which is even higher than in levitating droplets of Lee et al (S ≈ 4.1).[3] This can be explained in terms of the droplet size; our sessile microdroplets are thousand times smaller than that of Lee et al[3] (30-200 pL vs 1-8 µL). Indeed, smaller droplets can achieve deeper supersaturation levels due to the kinetic and thermodynamic confinement effects.[13]

Finally, by applying a fitting method (inhomogeneous Poisson distribution coupled with classical nucleation theory)[8], we obtained the pre-exponential factor (log A) and the effective interfacial energy ($\gamma_{eff}$) as displayed in Table 1. To evaluate the reproducibility across the 3 independent trials, we again applied the Anderson Darling test, which consequently confirmed that the 3 datasets have statistically identical distribution at 95% confidence (p-value = 0.09)

**Table 1.** Fitted pre-exponential factor and interfacial energy of the metastable KDP Phase IV from the probability distribution of nucleation times in Figure 4.

|         | N  | log A (m$^{-3}$s$^{-1}$) | $\gamma_{eff}$ (mJ/m$^2$) |
|---------|----|--------------------------|---------------------------|
| Trial 1 | 68 | 36.7                     | 56.1                      |
| Trial 2 | 46 | 38.0                     | 57.2                      |
| Trial 3 | 50 | 43.9                     | 60.9                      |
| Mean    | -- | 39.5 (±8%)               | 58.1 (±4%)                |

Our estimated effective aqueous interfacial energy $\gamma_{eff}$ for the metastable KDP Phase IV based on the mean of 3 trials is 58 mJ/m$^2$. Alternatively, fitting using the combined dataset (N= 164) resulted in $\gamma_{eff}$ = 57 mJ/m$^2$ (which is essentially equivalent). To the best of our knowledge, this is the first experimental report of the aqueous interfacial energy of KDP Phase IV based on classical nucleation theory and its experimental solubility.

In general, metastable forms are expected to have lower $\gamma_{eff}$ in comparison with the stable form. To estimate the $\gamma_{eff}$ of the stable form, we employ the empirical correlation by Mersmann[7] which can be written as

$$\gamma_{SL} = K(c_{sol})^{2/3} \ln\left(\frac{c_{sol}}{c_{sat}}\right) \qquad (3)$$

with $\gamma_{SL}$ as the interfacial energy (between crystal and solution), K as an empirical constant (identical for both polymorphs), and $c_{sol}$ and $c_{sat}$ as the equilibrium solid-state and liquid-phase concentrations (in mol/m$^3$) respectively. For the stable Phase II at ambient conditions, $c_{sol} = 17179$ mol/m$^3$ and $c_{sat} = 1621.8$ mol/m$^3$ (see Table 1 of Ref[7]). For the metastable Phase IV, $c_{sol} = 17032$ mol/m$^3$ (from the ratio of solid-state densities). Applying ratio and proportion between the stable and metastable form, the empirical constant K is canceled out. This effectively removes the dependence of the model prediction on the accuracy of the empirical fitting parameter $K$ (originally defined as $0.414 k_b T\, N_A^{2/3}$ where $k_bT$ and $N_A$ correspond to thermal energy and Avogadro's number respectively)[7].

Using our measured solubility ratio of 1.8, the Mersmann correlation yields $\gamma_{stable}/\gamma_{metastable}$ = 1.35. Knowing that $\gamma_{metastable}$= 58 mJ/m², the effective interfacial energy barrier of the stable form (Phase II) is then predicted to be around 78 mJ/m².

With the knowledge of the interfacial energies $\gamma_{eff}$ of both the stable Phase II and metastable Phase IV, we can proceed to model the competitive nucleation of the two polymorphs using classical nucleation theory (equation 2). Given that the mass transfer properties of the supersaturated solution (i.e. diffusivity, viscosity, etc) are identical in the liquid phase for both polymorphs, we can suppose that the parameter $A$ (related to the mass transfer rate of the monomers towards the nucleus) is similar for both polymorphs, as in the analysis of Sato[14] and Deij et al.[15] Thus, for comparison, we can use the term (J/A) to describe the relative nucleation rate. Using $\gamma_{eff}$ = 58 mJ/m² for the metastable Phase IV (obtained from nucleation statistics) and $\gamma_{eff}$ = 78 mJ/m² for the stable Phase II (Mersmann correlation), the relative nucleation rates as a function of supersaturation ratio (with respect to both forms) are shown in **Figure 5**.

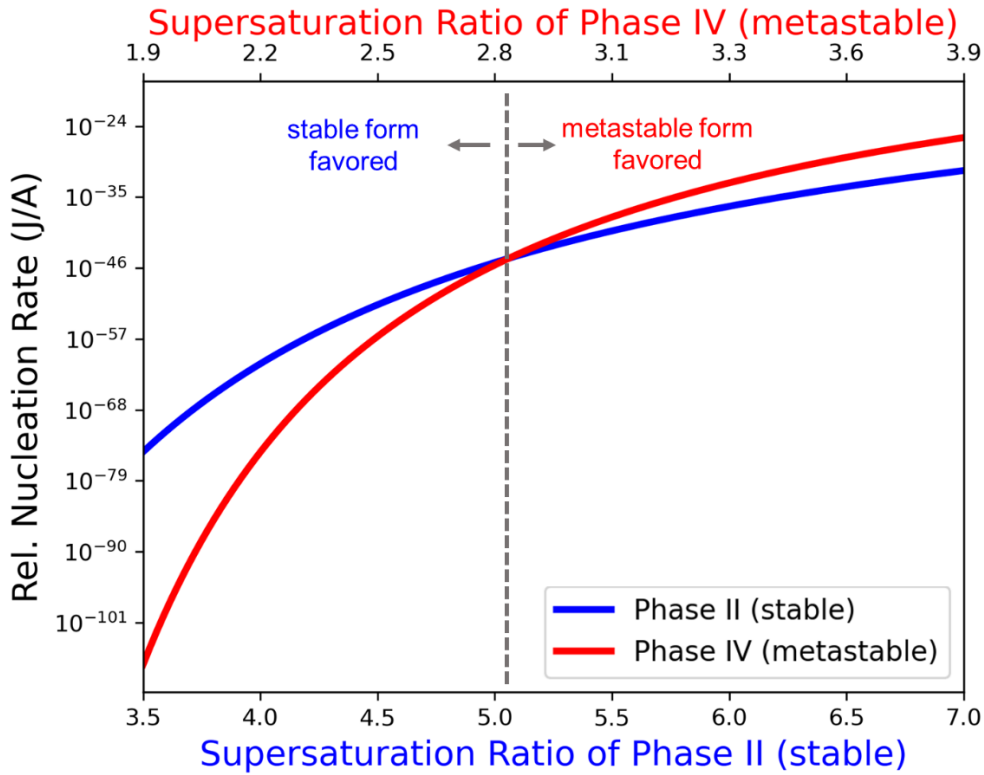

**Figure 5.** Relative nucleation rate ln(J/A) as a function of supersaturation ratio with respect to the stable phase (bottom axis) and the metastable phase (top axis).

We observe that the stable form is favored at low supersaturations while the metastable one at high supersaturations. This result is consistent with the experimental observation of Lee et al.[3] This phenomenon can be explained by the competition between thermodynamics and kinetics.[16] Although the metastable form has a lower nucleation barrier (kinetically favored), it experiences a lower supersaturation due to its high solubility (thermodynamically hindered). On the other hand, the stable form experiences a higher supersaturation (thermodynamically favored) but it needs to overcome a higher nucleation barrier (kinetically hindered). Thus, in the case of KDP, the stable Phase II dominates at lower concentrations because the highly soluble Phase IV is not sufficiently supersaturated. At higher concentrations, Phase IV dominates as it achieves sufficient supersaturation to overcome its lower nucleation barrier. Overall, our microfluidic experiments and modeling study reveal interesting insights on the competitive nucleation of KDP polymorphs.

## CONCLUSION

In this work, we demonstrate a rapid microfluidic approach to measure the aqueous solubility and interfacial energy of a metastable polymorph using KDP Phase IV as a model compound. Using our measured solubility, we estimated its interfacial energy from its nucleation time statistics. Our results reveal that the KDP Phase IV is about 1.8 times more soluble in water than the stable Phase II, and its estimated aqueous interfacial energy is around 58 mJ/m$^2$. Applying Mersmann correlation[7], the interfacial energy of the stable phase II is predicted to be around 78 mJ/m$^2$. With these interfacial energies, we used the classical nucleation theory to model the competitive nucleation of both polymorphs and the results show that the stable phase is favored at lower supersaturation while the metastable phase is favored at higher supersaturation, which agrees with our observations and experimental reports in the literature. Indeed, these findings reveal an interesting interplay between thermodynamics and kinetics in competitive polymorphic nucleation. Moreover, our results also highlight the advantage of the small volume (pL range) achieved in our microdroplet approach which permits access to unprecedentedly deep levels of supersaturation that would otherwise be unattainable with µL scale experiments.

Our results also open routes for future investigations. Knowing that the stable form is favored at low supersaturations, it would be interesting to develop microfluidic platforms that offer a wider range of accessible concentrations and volume . Furthermore, our results can be used for comparison with other experimental approaches. For instance, interfacial energies can be accessed by measuring the contact angle of the saturated solution with the crystal face.[17] Finally, the microfluidic approach presented herein can be used to study other polymorphic systems which are ubiquitous in the mineral processing and pharmaceutical industry.

## ACKNOWLEDGEMENT

RC acknowledges the financial support of ANR-FACET, ANR-19-CE08-0014-02.

Supplementary Material for:

# Microdroplet Approach for Measuring Aqueous Solubility and Nucleation Kinetics of a Metastable Polymorph: The case of KDP Phase IV

Ruel Cedeno[1], Romain Grossier[1], Nadine Candoni[1], Stéphane Veesler[1]*

[1]CNRS, Aix-Marseille Université, CINaM (Centre Interdisciplinaire de Nanosciences de Marseille), Campus de Luminy, Case 913, F-13288 Marseille Cedex 09, France

**S1. Evaporation Model**

**Table S1** Numerical values used as input in evaporation model[1] for aqueous KDP droplets, referenced at 25⁰C and 1 atm.

| Quantity | Symbol | Value | Unit |
|---|---|---|---|
| solubility of water in PDMS oil[1] | $c_s$ | 8.76 | mol/m³ |
| diffusivity of water in PDMS oil[1] | $D$ | 6.74 × 10⁻⁹ | m²s⁻¹ |
| supersaturation at matching time* | $S_{match}$ | 4.90 | - |
| coefficient of density change for KDP[2] | $b_1$ | 0.0738 | - |
| coefficient of water activity lowering for KDP** | $b_2$ | 0.130 | - |
| solubility of KDP in water[3] | $c_{eq}$ | 1.83 | mol/kg |
| molar mass of KDP[4] | $M_{NaCl}$ | 0.136 | kg/mol |
| diffusivity of KDP in water[5] | $D_i$ | 1.27×10⁻⁹ | m²/s |
| density of pure water[4] | $\rho_w$ | 997 | kg/m³ |

*measured by monitoring/interpolating the droplet volume (lateral view) as it optically disappears
**measured by fitting the evolution of droplet volume with the evaporation model

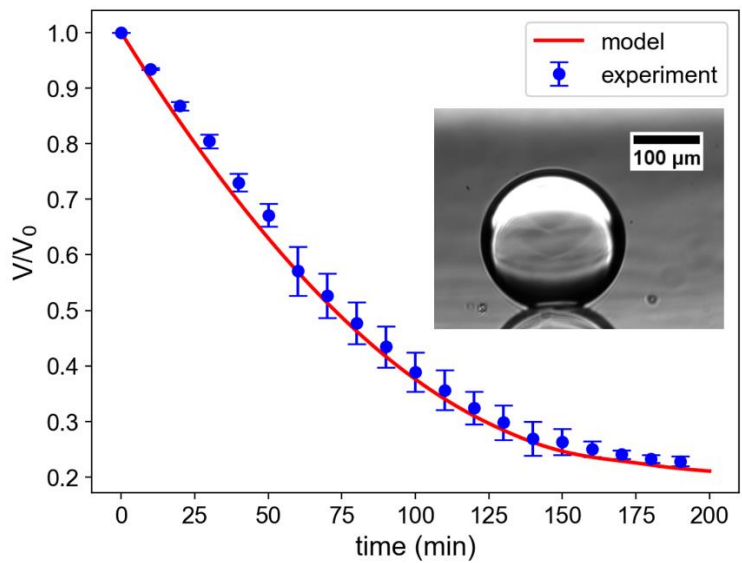

**Figure S1.** Evolution of relative droplet volume (V/V₀) as a function of time (3 replicates) taken from lateral images which were used to calibrate the evaporation model (for water activity dependence on supersaturation ratio).

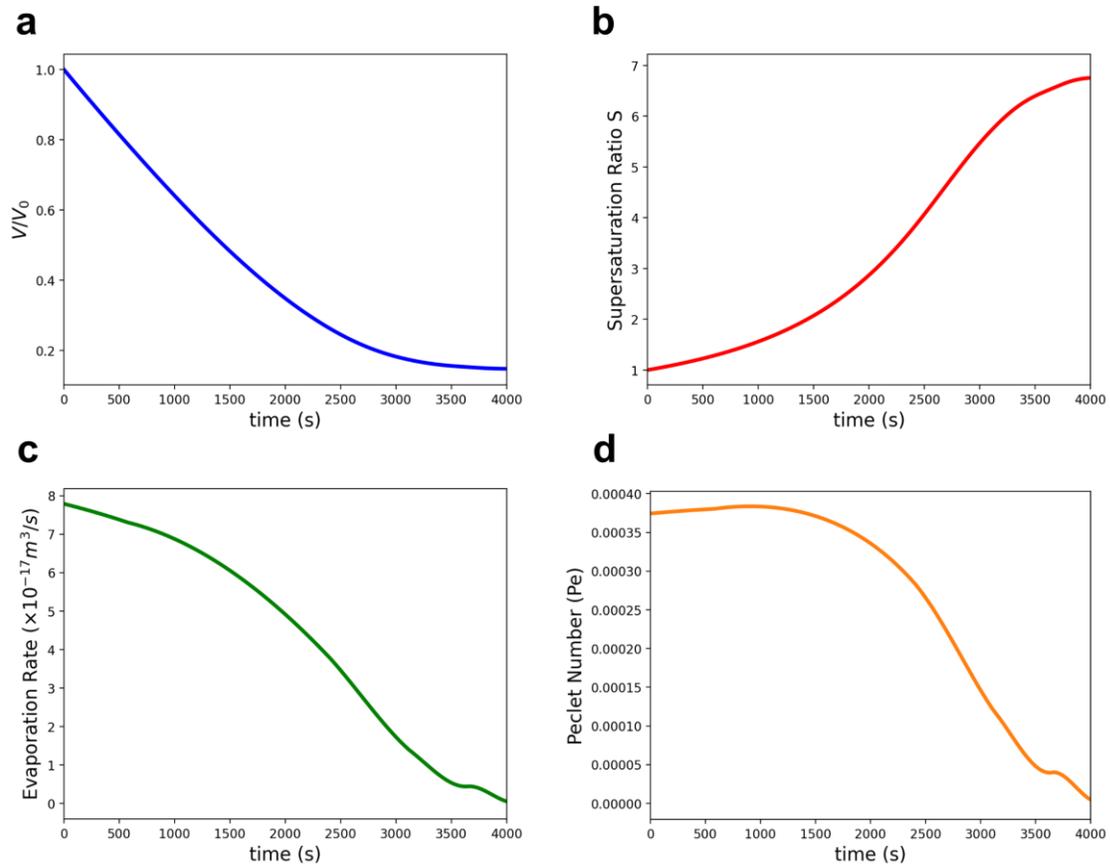

**Figure S2.** Model predictions for bottom-view arrays of microdroplets in terms of (a) relative volume $V/V_0$ (b) supersaturation ratio (c) evaporation rate (d) Peclet number[1, 6]. Pe < 1 suggests a uniform distribution of concentration within the droplet. The droplets were subjected under the following conditions: RH = 0.10, $S_0$ = 1, $V_0$ = 205 pL, $n_x$ = 3, $n_y$ = 1 where RH is the relative humidity, $S_0$ is the initial supersaturation ratio, $V_0$ is the initial volume, $n_x$ and $n_y$ are the effective number of intraline and interline interactions respectively. Note that we have demonstrated in Ref[7] (section S4.4) that the resulting τ distributions are not sensitive to the choice of $n_x$ and $n_y$.